\documentclass[]{JHEP3}   
 \preprint{  }

\usepackage{epsfig,multicol}

\title{Neutrino quasinormal modes of the Reissner-Nordstr\"om black hole}
\author{Jiliang Jing\thanks{Electronic address: jljing@hunnu.edu.cn}\\
 Institute of Physics and  Department of Physics, \\
Hunan Normal University,\\ Changsha, Hunan 410081, P. R. China }

\abstract{The neutrino quasinormal modes of the
Reissner-Nordstr\"om (RN) black hole are investigated using
continued fraction approach. We find, for large angular quantum
number, that the quasinormal frequencies become evenly spaced and
the spacing of the real part depends on the charge of the black
hole and that of the imaginary part is zero. We then find that the
quasinormal frequencies in the complex $\omega$ plane move
counterclockwise as the charge increases. They get a spiral-like
shape, moving out of their Schwarzschild value and ``looping in"
towards some limiting frequency as the charge tends to the
extremal value. The number of the spirals increases as the
overtone number increases but it decreases as the angular quantum
number increases. We also find that both the real and imaginary
parts are oscillatory functions of the charge, and the oscillation
becomes faster as the overtone number increases but it becomes
slower as the angular quantum number increases.}


\begin{document}

\section{introduction}

It is well-known that the quasinormal frequencies of a black hole
are defined as proper solutions of the perturbation equations
belonging to certain complex characteristic frequencies which
satisfy the boundary conditions appropriate for purely ingoing
waves at the event horizon and purely outgoing waves at infinity
\cite{Chand75}. The quasinormal modes (QNMs) are entirely fixed by
the structure of the background spacetime and are irrelevant of
the initial perturbations \cite{Chand75}. Thus, it is generally
believed that QNMs carry a unique footprint to directly identify
the existence of a black hole. Through the QNMs, one can extract
information about the physical parameters of the black
hole---mass, electric charge, and angular momentum---from the
gravitational wave signal by fitting the few lowest observed
quasinormal frequencies to those calculated from the perturbation
analysis. Meanwhile, the study of QNMs may lead to a deeper
understanding of the thermodynamic properties of black holes in
loop quantum gravity \cite{Hod} \cite{Dreyer}, as well as the QNMs
of anti-de Sitter black holes have a direction interpretation in
terms of the dual conformal field theory \cite{Maldacena}
\cite{Witten} \cite{Kalyana}. Therefore, much attention has been
devoted to the study of the quasinormal modes in the recent thirty
years \cite{Chan}-\cite{Konoplya1}.

Although QNMs of scalar, electromagnetic and gravitational
perturbations are studied extensively, the investigation of the
QNMs of the Dirac field is very limited \cite{Cho}\cite{Zhidenko}
\cite{Jing1}. Furthermore, in these papers the study of the Dirac
QNMs were limited to use WKB or P\"oshl-Teller potential
approximative methods. The WKB  and P\"oshl-Teller potential
approximative methods may be valuable tools in certain situations,
but in general results obtained from the approximative methods
should not be expected to be accurate. It is well known that the
Leaver's continued fraction technique is the best workhorse method
to compute highly damped modes and is very reliable. However, the
method can not be directly used to study the QNMs of the Dirac
fields if we take the standard wave equations.  The reason to
obstruct the study of the Dirac QNMs using continued fraction
technique is that for static black holes the standard wave
equation
\begin{eqnarray}\label{sta}
\left(\frac{d^2}{d r_*^2}+\omega^2\right)Z_{\pm}=V_{\pm}Z_{\pm}
\end{eqnarray}
possesses a special potential
 \begin{eqnarray}
V_{\pm}=\lambda^2\frac{ \Delta }{r^4}\pm \lambda \frac{d}{d
r_*}\frac{\sqrt{\Delta}}{r^2} \label{V00}
 \end{eqnarray}
which is the function of $\sqrt{\Delta}$ \cite{Cho}\cite{Jing1},
where $\Delta$ is a function related to the metric, say
$\Delta=r^2-2Mr+Q^2$ for the RN black hole. We can not investigate
the Dirac QNMs using the continued fraction approaches because we
have to expand the potential as a series at the event horizon but
the factor $\sqrt{\Delta}$ does not permit us to do that.

Recently, we \cite{Jing3} found that the wave function and
potential of the Dirac field can be expressed as new forms, and
the new wave function are related to the standard one in a simple
way \cite{Jing3}. Starting from the new wave function and
potential, we \cite{Jing3,Jing4} studied that the Dirac QNMs of
the Schwarzschild black hole using continued fraction
\cite{Leaver} and Hill-determinant approaches
\cite{Majumdar}\cite{Leaver2} and the Dirac QNMs of the
Schwarzschild-anti-de Sitter and Reissner-Nordstr\"om-anti-de
Sitter black holes using Horowitz-Hubeny approach \cite{Horowitz}.
Considering the RN black hole spacetime provides a better
background than the Schwarzschild geometry, in this paper we will
extend the study in Ref. \cite {Jing3} to the RN black hole, and
to see how the quasinormal frequencies depend on the charge of the
black hole.

The organization of this paper is as follows. In Sec. 2 the
decoupled neutrino equations and corresponding wave equation in
the RN spacetime are obtained using Newman-Penrose formalism. In
Sec. 3 the numerical approach to compute the neutrino QNMs is
introduced. In Sec. 4 the numerical results for the neutrino QNMs
in the RN black hole are presented. The last section is devoted to
a summary.

\section{Neutrino equations in the Reissner-Nordstr\"om spacetime}

The neutrino equations \cite{Page} can be expressed as
\begin{eqnarray}
   &&\sqrt{2}\nabla_{BB'}P^B=0, \nonumber \\
   &&\sqrt{2}\nabla_{BB'}Q^B=0,
\end{eqnarray}
where $\nabla_{BB'}$ is covariant differentiation, $P^B$ and $Q^B$
are the two-component spinors representing the wave functions, $
\bar{P}_{B'}$ is the complex conjugate of $P_{B}$. In the
Newman-Penrose formalism \cite{Newman} the equations become
\begin{eqnarray}\label{np}
   &&(D+\epsilon-\rho )P^0+
   (\bar{\delta}+\pi-\alpha )P^1=0,\nonumber \\
   &&(\triangle+\mu -\gamma )P^1+
   (\delta+\beta -\tau )P^0=0, \nonumber\\
   &&(D+\bar{\epsilon}-\bar{\rho} )\bar{Q}^{0'}+
   (\delta+\bar{\pi}-\bar{\alpha} )\bar{Q}^{1'}=0,\nonumber \\
  &&(\triangle+\bar{\mu} -\bar{\gamma} )\bar{Q}^{1'}+
   (\bar{\delta}+\bar{\beta} -\bar{\tau} )\bar{Q}^{0'}=0.
\end{eqnarray}
The null tetrad for the RN black hole can be taken as
\begin{eqnarray}
  &&l^\mu=(\frac{r^2}{\Delta}, ~1, ~0, ~0 ), \nonumber \\
  &&n^\mu=\frac{1}{2}(1, ~-\frac{\Delta}{r^2}, ~0, ~0)\nonumber \\
  &&m^\mu=\frac{1}{\sqrt{2} r}\left(0, ~0, ~1, \frac{i}{sin\theta}\right),
\end{eqnarray}
with
\begin{eqnarray}
\Delta=r^2-2M r+Q^2=(r-r_+)(r-r_-),
\end{eqnarray}
where $M$ and $Q$ are the mass and charge of the RN black hole,
$r_{+}=M+ \sqrt{M^2-Q^2}$ and $r_{-}=M- \sqrt{M^2-Q^2}$ represent
the event and inner horizons. The wave functions can be expressed
as
\begin{eqnarray}
&&P^0=\frac{1}{r}{\mathbb{R}}_{-1/2}(r)S_{-1/2}(\theta)
e^{-i(\omega t-\bar{m}\varphi)}, \nonumber \\
&&P^1={\mathbb{R}}_{+1/2}(r)S_{+1/2}(\theta)
e^{-i(\omega t-\bar{m}\varphi)}, \nonumber \\
&&\bar{Q}^{1'}={\mathbb{R}}_{+1/2}(r)S_{-1/2}(\theta)
e^{-i(\omega t-\bar{m}\varphi)}, \nonumber \\
&&\bar{Q}^{0'}=-\frac{1}{r}{\mathbb{R}}_{-1/2}(r)S_{+1/2}(\theta)e^{-i(\omega
t-\bar{m}\varphi)},
\end{eqnarray}
where $\omega$ and $\bar{m}$ are the energy and angular momentum
of the neutrino particle. After the tedious calculation Eq.
(\ref{np}) can be simplified as
\begin{eqnarray}\label{dd2}
&&\sqrt{\Delta}{\mathcal{D}}_0 {\mathbb{R}}_{-1/2}=
\lambda\sqrt{\Delta} {\mathbb{R}}_{+1/2}, \\
\label{dd3}&&\sqrt{\Delta}{\mathcal{D}}_0^{\dag}
(\sqrt{\Delta}{\mathbb{R}}_{+1/2})=\lambda {\mathbb{R}}_{-1/2},\\
&&{\mathcal{L}}_{1/2} S_{+1/2}=-\lambda S_{-1/2}, \label{aa1}\\
&&{\mathcal{L}}_{1/2}^{\dag} S_{-1/2}=\lambda S_{+1/2},\label{aa2}
\end{eqnarray}
with
 \begin{eqnarray}
 &&{\mathcal{D}}_n=\frac{\partial}{\partial r}-\frac{i r^2 \omega}
 {\Delta}+\frac{n}{\Delta}\frac{d \Delta}{d r},\nonumber \\
 &&{\mathcal{D}}^{\dag}_n=\frac{\partial}{\partial r}+\frac{i r^2 \omega}
 {\Delta}+\frac{n}{\Delta}\frac{d \Delta}{d r},\nonumber \\
 &&{\mathcal{L}}_n=\frac{\partial}{\partial \theta}
 +\frac{m}{\sin \theta }
 +n\cot \theta,\nonumber \\
 &&{\mathcal{L}}^{\dag}_n=\frac{\partial}{\partial \theta}
 -\frac{m}{\sin \theta }
 +n\cot \theta,\label{ld}
 \end{eqnarray}
where $\lambda=\pm \left(l+\frac{1}{2}\right)$ ( $l$ is the
quantum number characterizing the angular distribution). We will
focus our attention on the massless Dirac field in this paper.
Therefore, we can eliminate ${\mathbb{R}}_{-1/2}$ (or
$\sqrt{\Delta}{\mathbb{R}}_{+1/2}$) from Eqs. (\ref{dd2}) and
(\ref{dd3}) to obtain a radial decoupled Dirac equation for
$\sqrt{\Delta} {\mathbb{R}}_{+1/2}$ (or ${\mathbb{R}}_{-1/2}$).
Then, introducing an usual tortoise coordinate
 \begin{eqnarray}
dr_*=\frac{r^2}{\Delta} dr,
 \end{eqnarray}
 and resolving the equation in the
form
 \begin{eqnarray}
 {\mathbb{R}}_{s}=\frac{\Delta^{-s/2}}{r} \Psi_s,
 \end{eqnarray}
we obtain the wave equation
\begin{eqnarray}\label{wave}
\frac{d^2 \Psi_s }{d r_*^2}+(\omega ^2-V_s )\Psi_s =0,
\end{eqnarray}
with
\begin{eqnarray}\label{Poten}
V_s=-\frac{\Delta}{4 r^2}\frac{d}{d
r}\left[r^2\frac{d}{dr}\left(\frac{\Delta}{r^4}\right)\right]+\frac{s^2
r^4}{4}\left[\frac{d}{d
r}\left(\frac{\Delta}{r^4}\right)\right]^2+is \omega r^2\frac{d}{d
r}\left(\frac{\Delta}{r^4}\right)+\frac{\lambda^2 \Delta}{r^4}.
 \end{eqnarray}
We will study the neutrino quasinormal frequencies of the RN black
hole using Eqs. (\ref{wave}) and (\ref{Poten}) and with the help
of the continued fraction approach.

\section{Numerical Approaches}

For the RN black hole, the QNMs are defined to be modes with
purely ingoing waves at the event horizon and purely outgoing
waves at infinity \cite{Chand75}. Then, the boundary conditions on
wave function $\Psi_s$ at the horizon $(r=r_+)$ and infinity
$(r\rightarrow +\infty)$ can be expressed as
 \begin{eqnarray}
 \label{Bon}
\Psi_s  \sim \left\{
\begin{array}{ll} (r-r_+)^{-\frac{s}{2}-\frac{i\omega}{2\kappa_{+}}} &
~~~~r\rightarrow r_+, \\
     r^{-s+i\omega}e^{i\omega r} & ~~~~     r\rightarrow +\infty,
\end{array} \right.
 \end{eqnarray}
where $\kappa_+=\frac{r_+-r_-}{2r^2_+}$ is the surface gravity on
the event horizon $r_+$.

A solution to Eq. (\ref{wave}) that has the desired behavior at
the boundary can be written in the form
 \begin{eqnarray}\label{expand}
 \Psi_s=r(r-r_+)^{-\frac{s}{2}
 -\frac{i\omega}{2\kappa_+}}(r-r_-)^{-1-\frac{s}{2}+2i\omega+
 \frac{i\omega}{2\kappa_-}}e^{i\omega (r-r_-)}\sum_{m=0}^{\infty}
 a_m\left(\frac{r-r_+}{r-r_-}\right)^m,
 \end{eqnarray}
where $\kappa_-=\frac{r_+-r_-}{2r^2_-}$ is the surface gravity on
the inner horizon $r_-$. If we take $r_++r_-=1$ and $b=r_+-r_-$,
the sequence of the expansion coefficients $\{a_m: m=1,2,....\}$
is determined by a three-term recurrence relation staring with
$a_0=1$:
 \begin{eqnarray} \label{rec}
 &&\alpha_0 a_1+\beta_0 a_0=0, \nonumber \\
 &&\alpha_m a_{m+1}+\beta_m a_m+\gamma_m a_{m-1}=0,~~~m=1,2,....
 \end{eqnarray}
The recurrence coefficient $\alpha_m$, $\beta_m$ and $\gamma_m$
are given in terms of $m$ and the black hole parameters by
\begin{eqnarray}
 &&\alpha_m=m^2+(C_0+1)m+C_0, \nonumber \\
 &&\beta_m=-2m^2+(C_1+2)m+C_3, \nonumber  \\
 &&\gamma_m=m^2+(C_2-3)m+C_4-C_2+2,
 \end{eqnarray}
and the intermediate constants $C_m$ are defined by
\begin{eqnarray}
 &&C_0=1-s-i\omega-\frac{i\omega(r_+^2+r_-^2)}{b}, \nonumber \\
 &&C_1=-4+2i\omega(2+b)+\frac{2i\omega(r_+^2+r_-^2)}{b},  \nonumber  \\
 &&C_2=s+3-3i\omega-\frac{i\omega(r_+^2+r_-^2)}{b}, \nonumber \\
 &&C_3=\omega^2(4+2b-4r_+r_-)-s-1+(2+b)i\omega-\lambda^2+(2\omega+i)
 \frac{\omega(r_+^2+r_-^2)}{b}, \nonumber \\
 &&C_4=s+1-2\omega^2-(2s+3)i\omega-(2\omega+i)\frac{\omega(r_+^2+r_-^2)}{b}.
 \end{eqnarray}
It is interesting to note that the three-term recursion relation
is obtained form the wave equation (\ref{wave}) directly. The
calculation is simpler than other cases. For example, the
coefficients of the expansion for the electromagnetic and
gravitational fields in the RN black hole are determined by a
four-term recursion relation which should be reduced to a
three-term relation using a Gaussian eliminated step (see
\cite{Leaver2} for details).

The series in (\ref{expand}) converges and the $r=+\infty$
boundary condition (\ref{Bon}) is satisfied if, for a given $s$
and $\lambda$, the frequency $\omega$ is a root of the continued
fraction equation
\begin{eqnarray}\label{ann}
 \left[\beta_m-\frac{\alpha_{m-1}\gamma_m}{\beta_{m-1}-}
 \frac{\alpha_{m-2}\gamma_{m-1}}{\beta_{m-2}-}...
 \frac{\alpha_0\gamma_1}{\beta_0}\right]=
 \left[\frac{\alpha_m\gamma_{m+1}}{\beta_{m+1}-}
 \frac{\alpha_{m+1}\gamma_{m+2}}{\beta_{m+2}-}
 \frac{\alpha_{m+2}\gamma_{m+3}}{\beta_{m+3}-}...\right],
 ~~(m=1,2...).
 \end{eqnarray}
This leads to a simple method to find quasinormal frequencies---
defining a function which returns the value of the continued
fraction for an initial guess at the frequency, and then use a
root finding routine to find the zeros of this function in the
complex plane. The frequency for which happens is a quasinormal
frequency. The $n$th quasinormal frequency is usually found to be
the most stable root of the $n$th inversion \cite{Leaver}.

\section{Neutrino Quasinormal Modes of the Reissner-Nordstr\"om black hole}

  In this section we will present the numerical results of the neutrino
quasinormal frequencies of the RN black hole obtained by using the
numerical approach just outlined in the previous section. The
results will be organized into three subsections: the fundamental
quasinormal frequencies, the QNMs for large angular quantum number
and highly damped modes.

\subsection{Fundamental quasinormal frequencies}

We find that the change of the neutrino quasinormal frequencies
becomes faster as the charge increases. Therefore, as we approach
the extremal value $Q=1/2$, the convergence of the continued
fraction method slower, and the required computing time gets
longer. The fundamental neutrino quasinormal frequencies of the RN
black hole for $Q=0$ to $Q=0.499$ and $\lambda=1$ are listed in
the table I. We know from the table that the real part of the
fundamental quasinormal frequency increases as $Q$ increases, but
the imaginary part decreases first and then increases as $Q$
increases (we should note that the magnitude of the imaginary part
increases first and then decreases). It is shown that the
intermediate decay of the Dirac perturbation around the RN black
hole depends on the charge $Q$. The oscillating frequency of this
decay which corresponds to $\omega_{R}$ increases as $Q$
increases, and the amplitude per unit time $\sim e^{\omega_{I}}$
increases first and then decreases. We should point out that this
property is true of all fundamental quasinormal frequencies with
different angular quantum number.

\TABLE{ \caption{\label{table1} The fundamental neutrino
quasinormal frequencies of the RN black hole for $Q=0$ to
$Q=0.499$ and $\lambda=1$. As charge $Q$ increases, the real part
of the fundamental quasinormal frequency increases, but the
imaginary part decreases first and then increases.}
\begin{tabular}{c|c|c|c|c|c}
 \hline \hline
 ~~$ Q $ ~~ & $\omega$  & ~~$Q$~~   & $\omega$ &~~$Q$ ~~ &
 $\omega$   \\
\hline
0.000 & ~ 0.365926 -0.193965i ~& 0.410 &~ 0.427193 -0.196008i ~& 0.460 & 0.452381 -0.190363i \\
0.050 &  0.366591 -0.194066i & 0.415 & 0.429341 -0.195734i & 0.465 & 0.455387 -0.189224i \\
0.100 &  0.368615 -0.194366i & 0.420 & 0.431562 -0.195418i & 0.470 & 0.458469 -0.187904i \\
0.150 &  0.372096 -0.194851i & 0.425 & 0.433862 -0.195052i & 0.475 & 0.461608 -0.186375i \\
0.200 &  0.377208 -0.195488i & 0.430 & 0.436242 -0.19463i  & 0.480 & 0.464769 -0.184605i \\
0.250 &  0.384241 -0.196213i & 0.435 & 0.438707 -0.194144i & 0.485 & 0.467892 -0.182571i  \\
0.300 &  0.393651 -0.196891i & 0.440 & 0.441258 -0.193585i & 0.490 & 0.47089 -0.180287i \\
0.350 &  0.406184 -0.19722i  & 0.445 & 0.4439 -0.192942i   & 0.495 & 0.473689 -0.177851i  \\
0.400 &  0.423108 -0.196443i & 0.450 & 0.446635 -0.192202i & 0.497 & 0.474767 -0.176866i \\
0.405 & 0.425116 -0.196243i & 0.455 & 0.449462 -0.191348i & 0.499 & 0.475833 -0.175873i \\
 \hline \hline
\end{tabular}}

\subsection{QNMs for large angular quantum number}

The quasinormal frequencies for $\lambda=1$ to $\lambda=40$ and
$n=0,~1,~2$ are described by Fig. \ref{fig1} which shows that
$\Delta \omega=\omega_{\lambda+1}-\omega_{\lambda}$ as a function
of $\lambda$. From the figure we know that, for the cases of
$n=0$, $n=1$  and $n=2$, the quasinormal frequencies become evenly
spaced for large $\lambda$ and the spacing is given by
\begin{eqnarray}
&&\Delta \omega(Q=0.0)=0.38490-0.0000i. \nonumber \\
&&\Delta \omega(Q=0.1)=0.38750-0.0000i. \nonumber \\
&&\Delta \omega(Q=0.2)=0.39581-0.0000i. \nonumber \\
&&\Delta \omega(Q=0.3)=0.41170-0.0000i.
\end{eqnarray}
That is to say, the spacing of the real part is related to the
charge $Q$ and the spacing of the imaginary part becomes zero for
large $\lambda$.

\FIGURE{\includegraphics[scale=0.7]{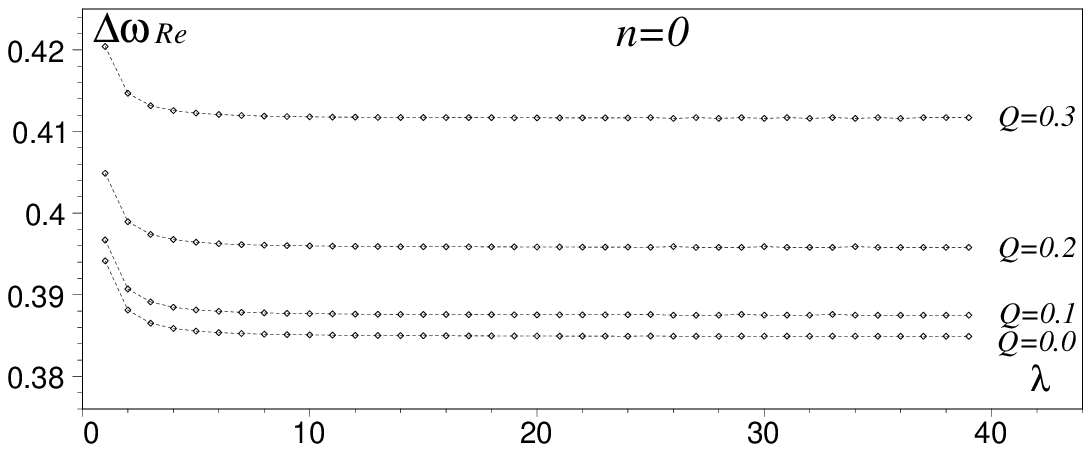}\hspace*{0.2cm}
\includegraphics[scale=0.7]{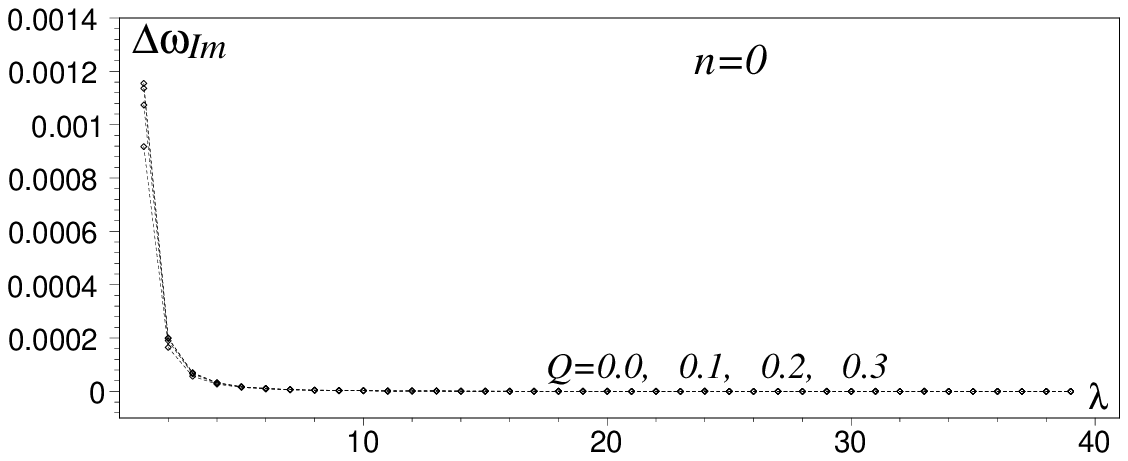} \vspace*{0.3cm} \\
\includegraphics[scale=0.7]{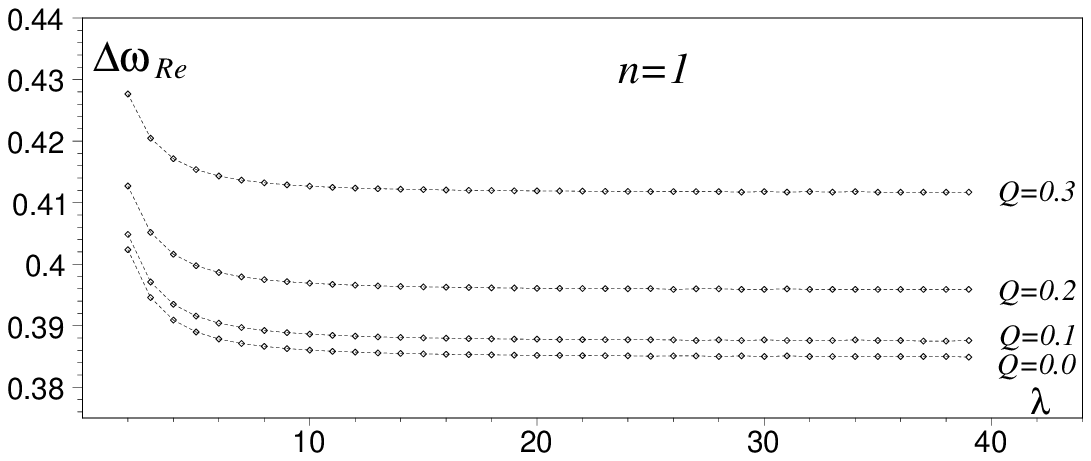}\hspace*{0.5cm}
\includegraphics[scale=0.7]{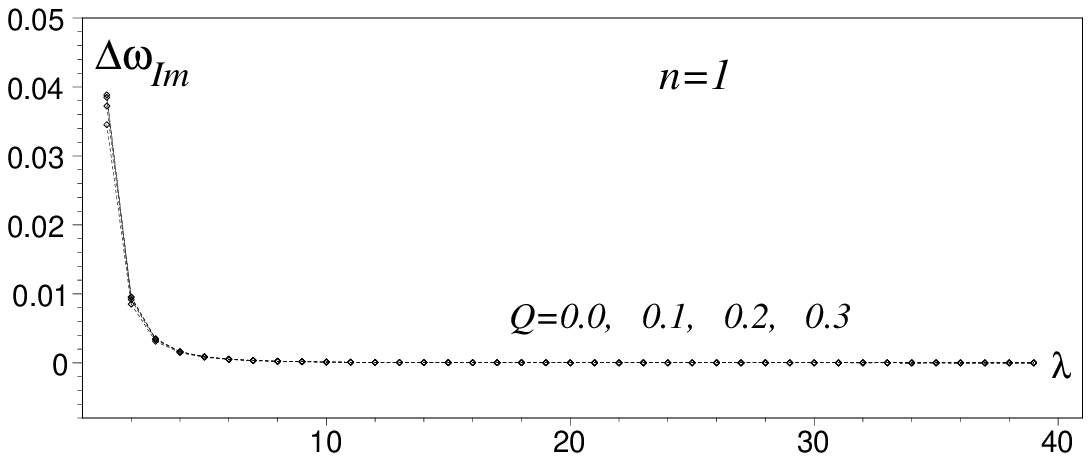} \vspace*{0.3cm} \\
\includegraphics[scale=0.7]{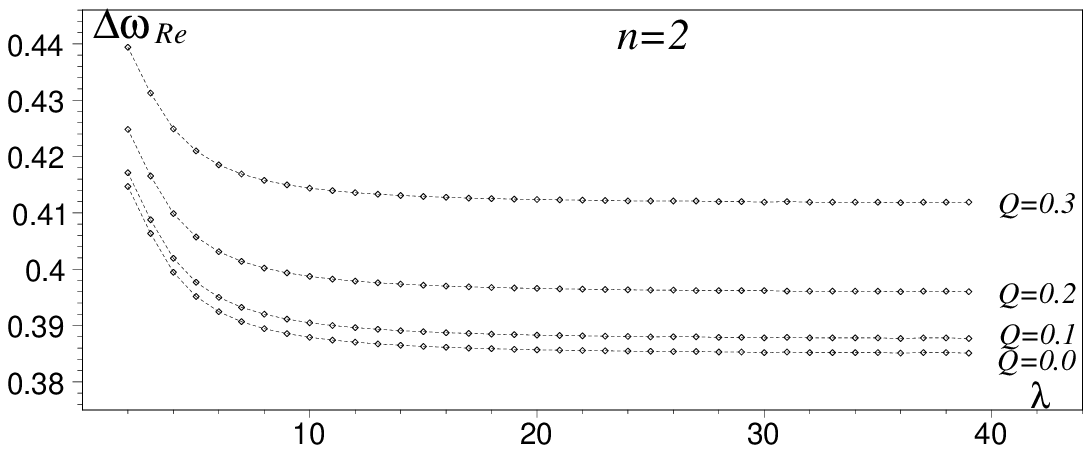}  \hspace*{0.5cm}
\includegraphics[scale=0.7]{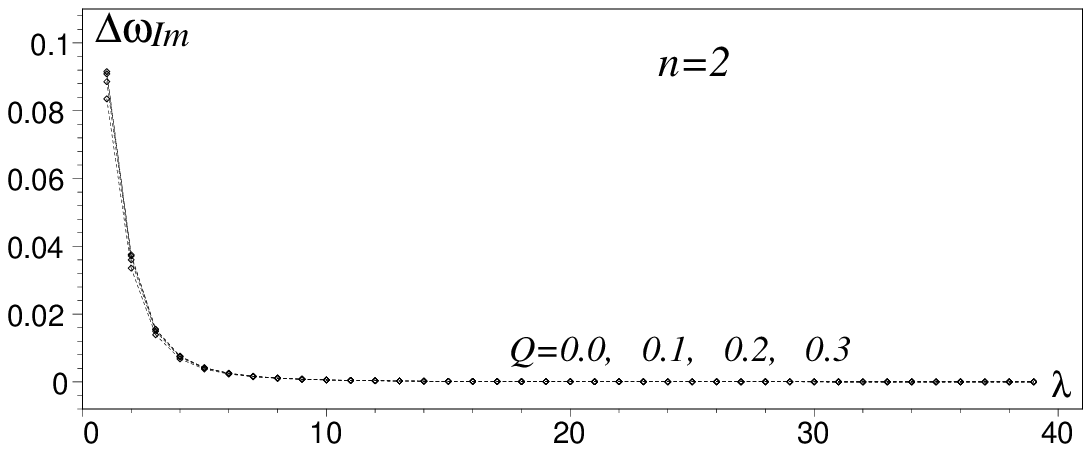}
\caption{\label{fig1} The spacing $\Delta
\omega=\omega_{\lambda+1}-\omega_{\lambda}$ as the functions of
$\lambda$ for $n=0,~1,~2$ quasinormal frequencies. The left three
figures are drawn for $\Delta \omega _{Re}
=Re(\omega_{\lambda+1})-Re(\omega_{\lambda})$ which show that the
spacing of the real part is related to the charge $Q$ for large
$\lambda$. The right three figures are for $\Delta
\omega_{Im}=Im(\omega_{\lambda+1})-Im(\omega_{\lambda})$ which
show that the spacing of the imaginary part becomes zero for large
$\lambda$.}}

\subsection{Highly damped modes}

 The Fig. \ref{fig2} is drawn for $n=0$ to $n=3$ and
$\lambda=1$, and Fig. \ref{fig3} is for $n=5,6,7,8,10$ and
$\lambda=2$. In these figures the quasinormal frequency
trajectories are formed only by points which are quasinormal
frequencies at increments in $Q$ of 0.01 for $Q=0$ to $Q\sim 0.40$
and at increments in $Q$ of 0.001 for $Q\sim 0.40$ to $Q=0.495$.
We now proceed to discussed the figures in more detail.

The left columns in the Figs. \ref{fig2} and \ref{fig3} describe
the behavior of quasinormal frequencies in the complex $\omega$
plane which show that the frequencies generally move
counterclockwise as the charge increases.  They get a spiral-like
shape, moving out of their Schwarzschild value ($Q=0$) and
``looping in" towards some limiting frequency as the charge $Q$
tends to the extremal value, $Q=1/2$. For a given $\lambda$, we
observe that  the number of spirals increases as the overtone
number increases. However, for a given overtone number $n$,
increasing $\lambda$  has the effect of ``unwinding" the spirals,
as we see in the two figures that the spiral begins at $n=2$ for
$\lambda=1$ but it starts at $n=6$ for $\lambda=2$.

The second and last columns in the Figs. \ref{fig2} and \ref{fig3}
illustrate that the real and imaginary parts of the quasinormal
frequencies are the functions of the charge $Q$. We know from
these figures that both the real and imaginary parts of the
frequencies are oscillatory functions of the charge. The
oscillation starts earlier and earlier as the overtone number $n$
grows for a fixed $\lambda$, but it begins later and later as the
angular quantum number increases for a fixed $n$. Meanwhile, the
oscillation becomes faster as the overtone number increases for a
given $\lambda$, but it becomes slower as $\lambda$ increases for
a given $n$.

\FIGURE{
\includegraphics[scale=0.6]{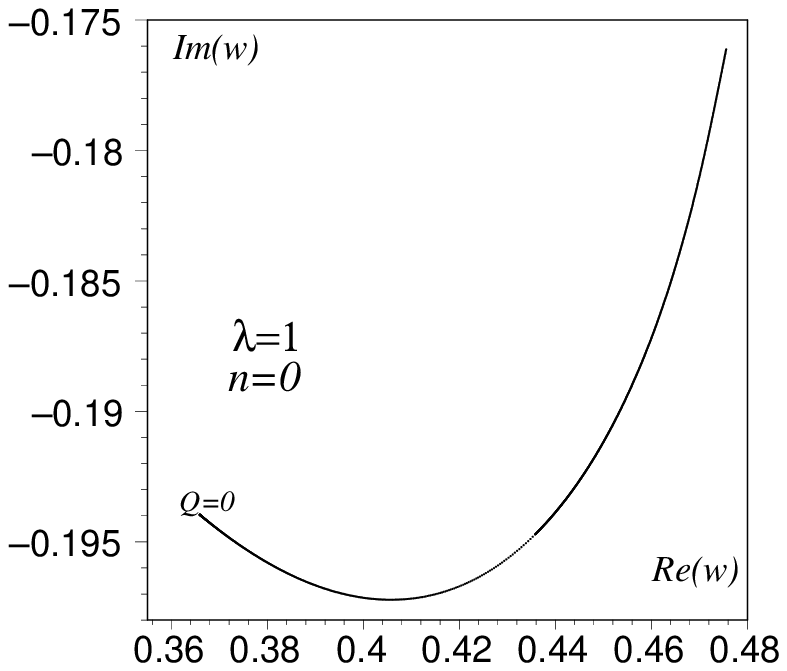}\hspace*{0.2cm}
\includegraphics[scale=0.6]{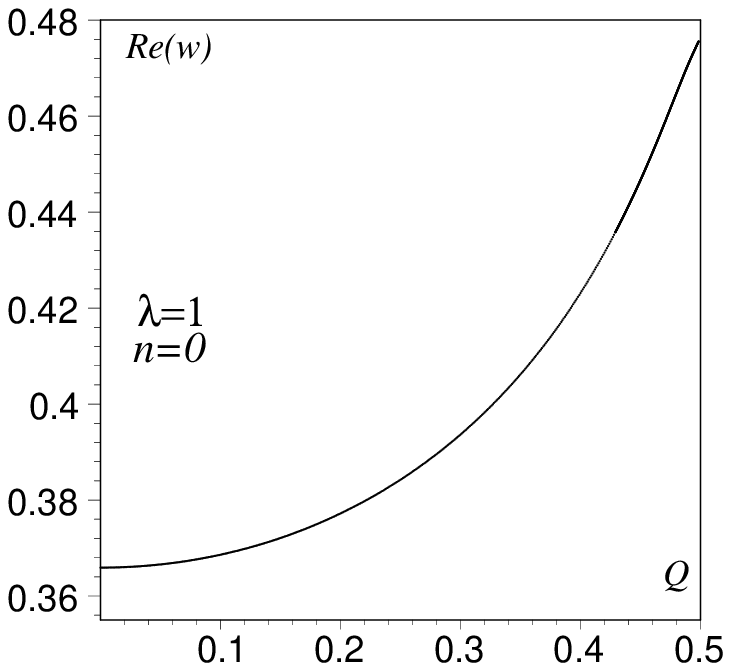} \hspace*{0.4cm}
\includegraphics[scale=0.6]{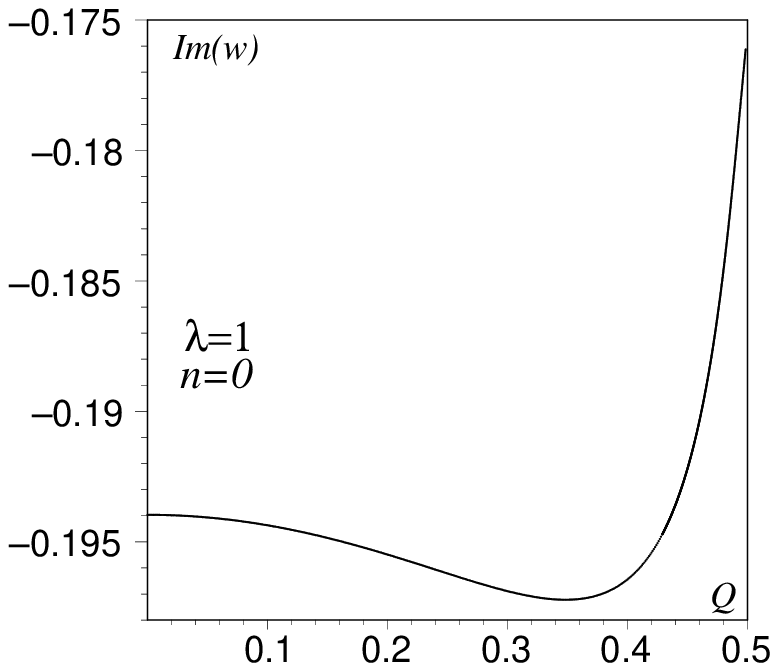} \vspace*{0.2cm} \\
\includegraphics[scale=0.6]{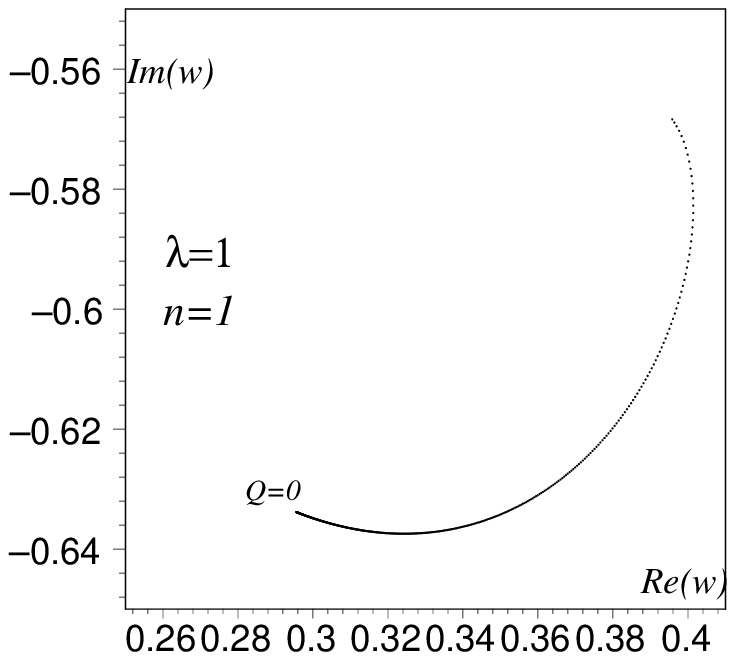}\hspace*{0.5cm}
\includegraphics[scale=0.6]{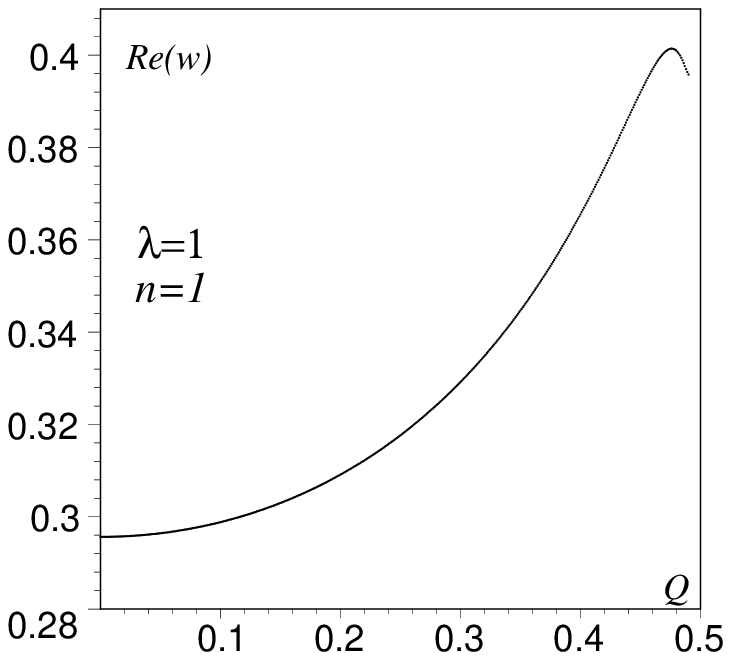} \hspace*{0.5cm}
\includegraphics[scale=0.6]{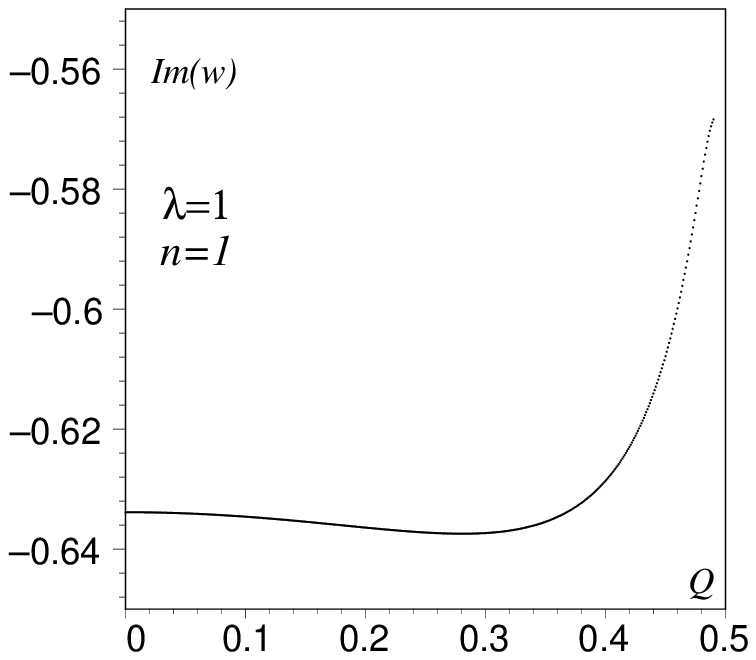}\vspace*{0.2cm} \\
\includegraphics[scale=0.6]{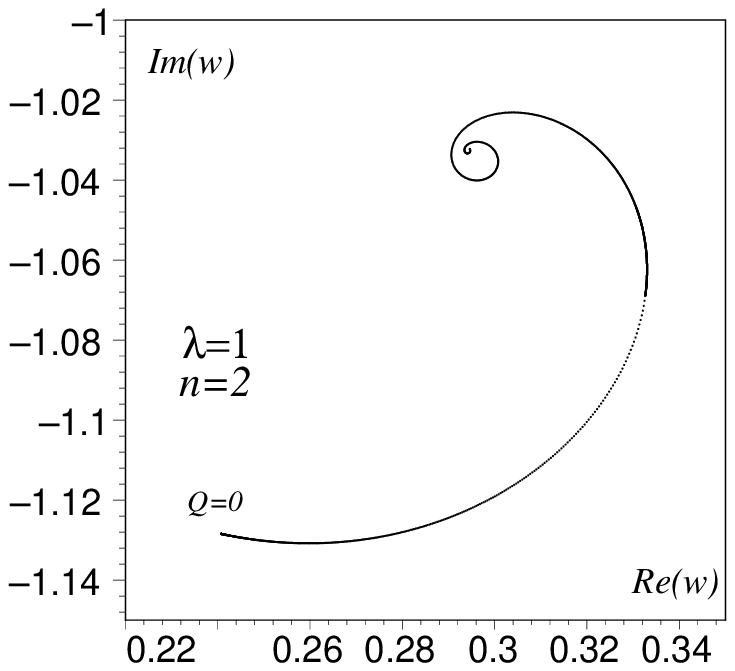} \hspace*{0.5cm}
\includegraphics[scale=0.6]{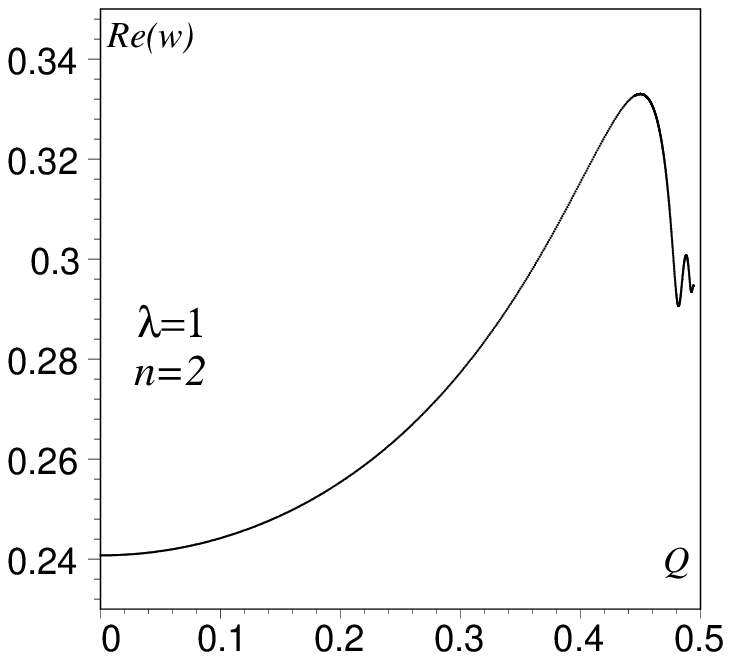} \hspace*{0.5cm}
\includegraphics[scale=0.6]{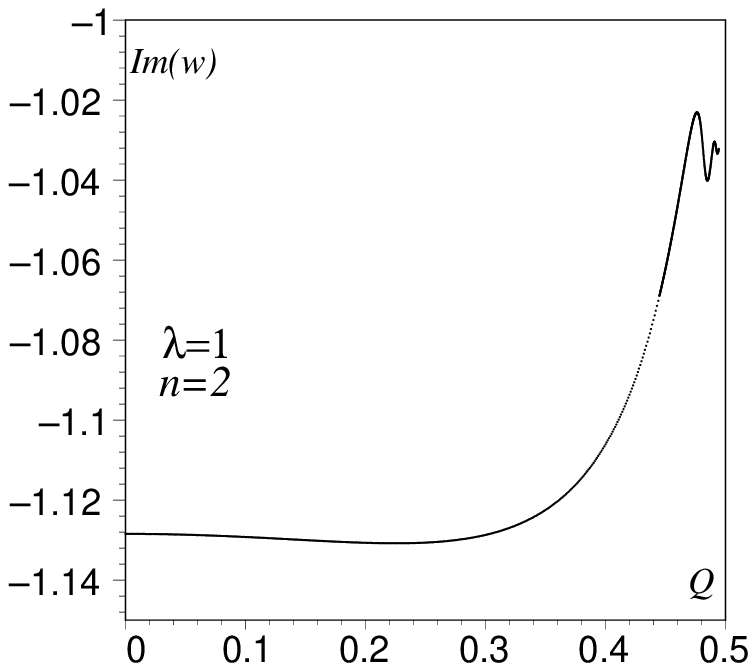} \vspace*{0.2cm}\\
\includegraphics[scale=0.6]{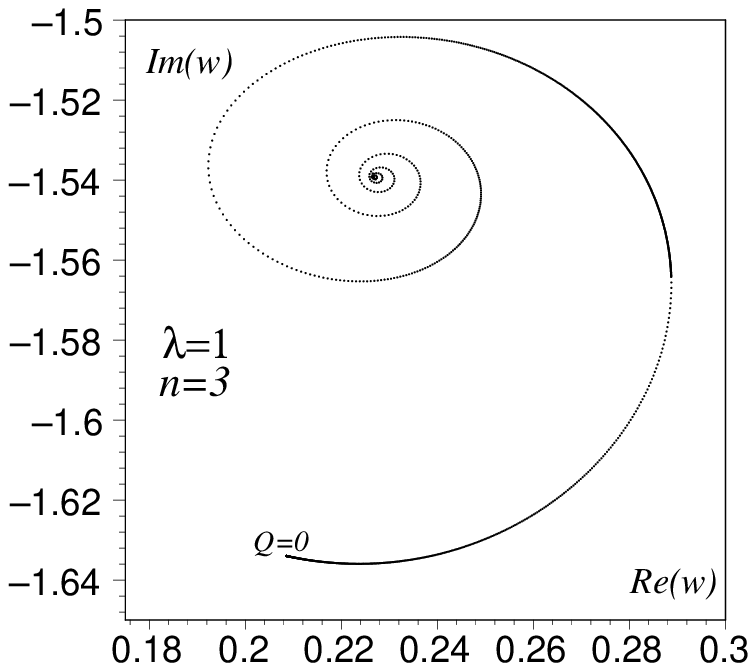} \hspace*{0.5cm}
\includegraphics[scale=0.6]{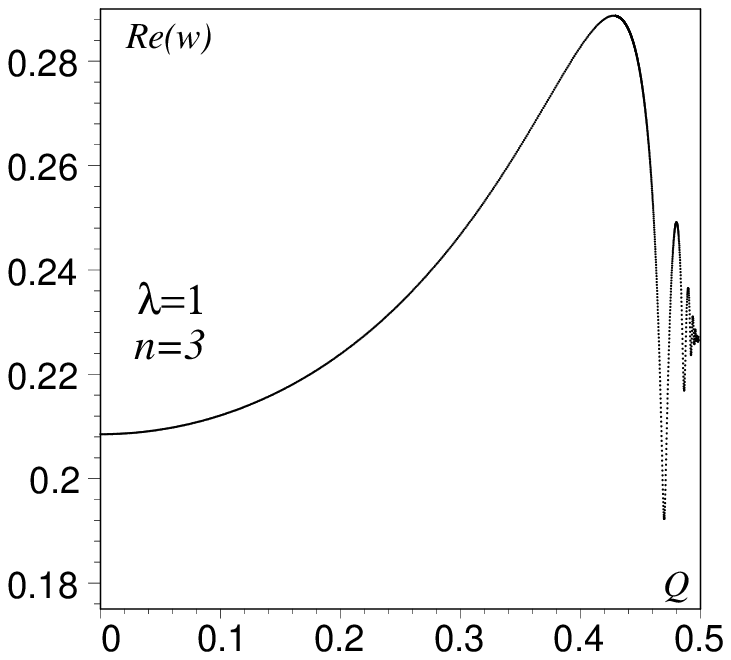} \hspace*{0.5cm}
\includegraphics[scale=0.6]{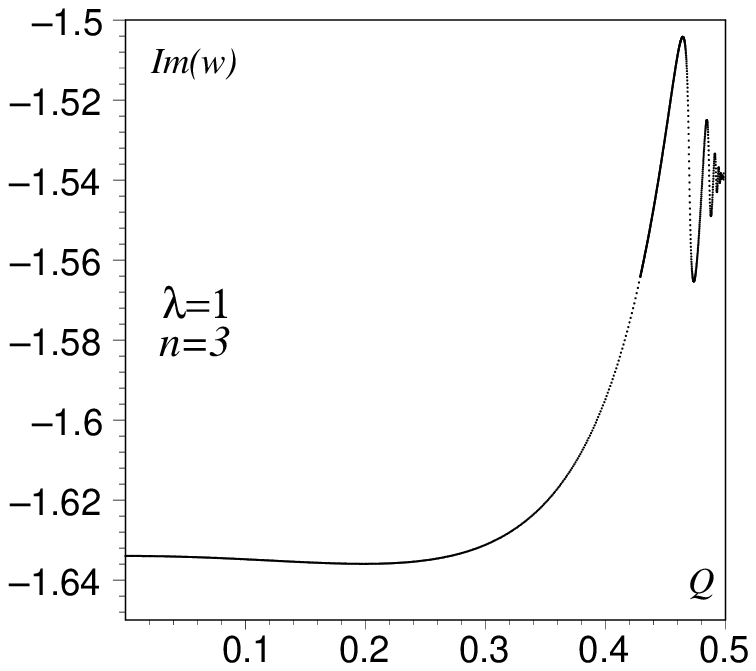} \vspace*{0.2cm}
\caption{\label{fig2} The left 4 figures describe the behavior of
quasinormal frequencies for $n=0$ to $n=3$ and $\lambda=1$ in the
complex $\omega$ plane which show that the frequencies generally
move counterclockwise as the charge is increased and the number of
spirals increases as the overtone number increases. The others
graph $Re(\omega)$ and $Im(\omega)$ of the quasinormal frequencies
versus charge $Q$ which tell us that both the real and imaginary
parts are oscillatory functions of charge and the oscillations
become faster as the overtone number increases.}}

\FIGURE{
\includegraphics[scale=0.6]{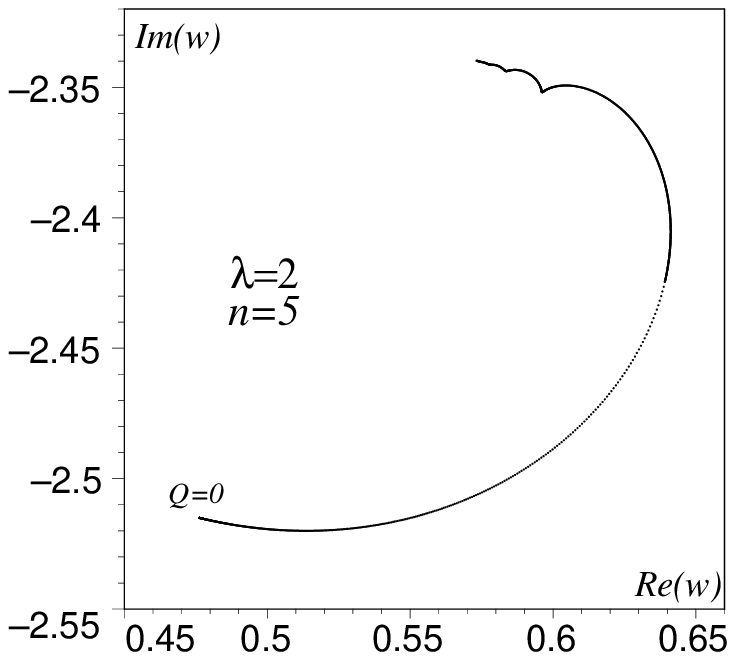}\hspace*{0.8cm}
\includegraphics[scale=0.6]{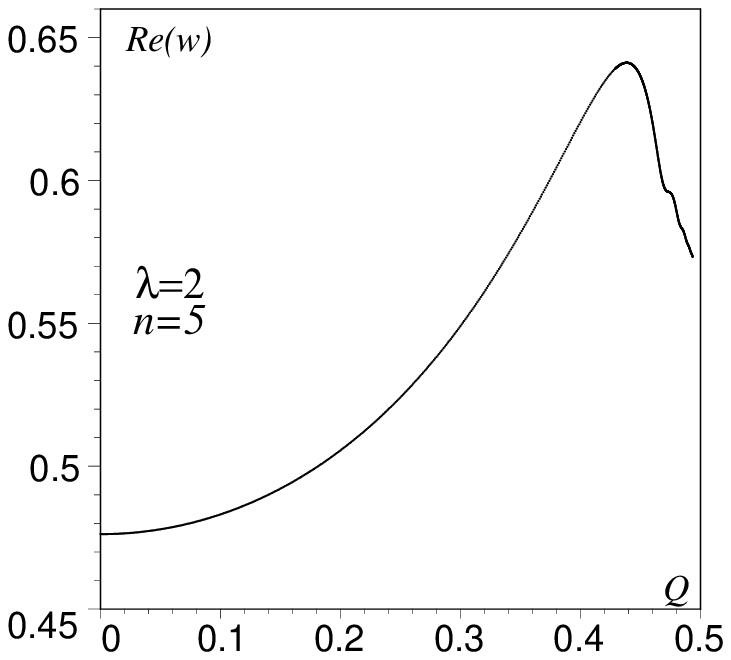} \hspace*{0.5cm}
\includegraphics[scale=0.6]{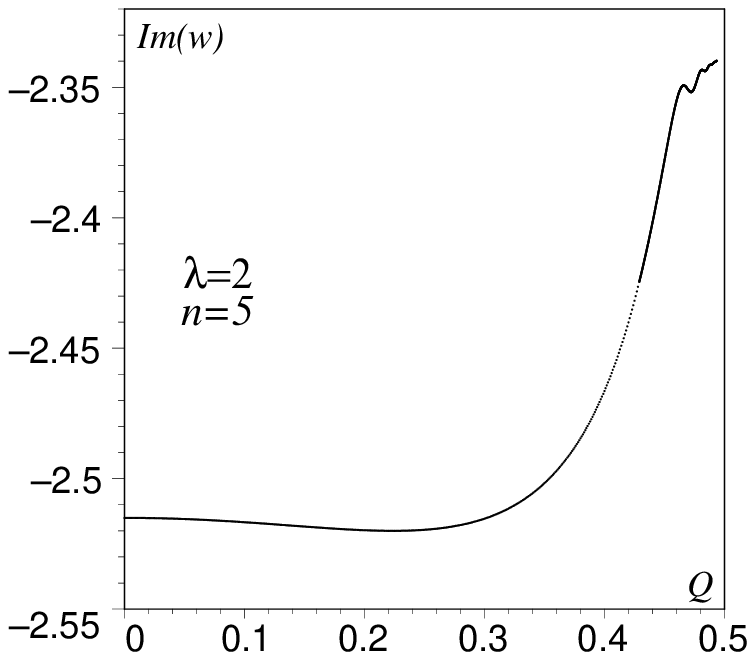}  \vspace*{0.2cm} \\
\includegraphics[scale=0.6]{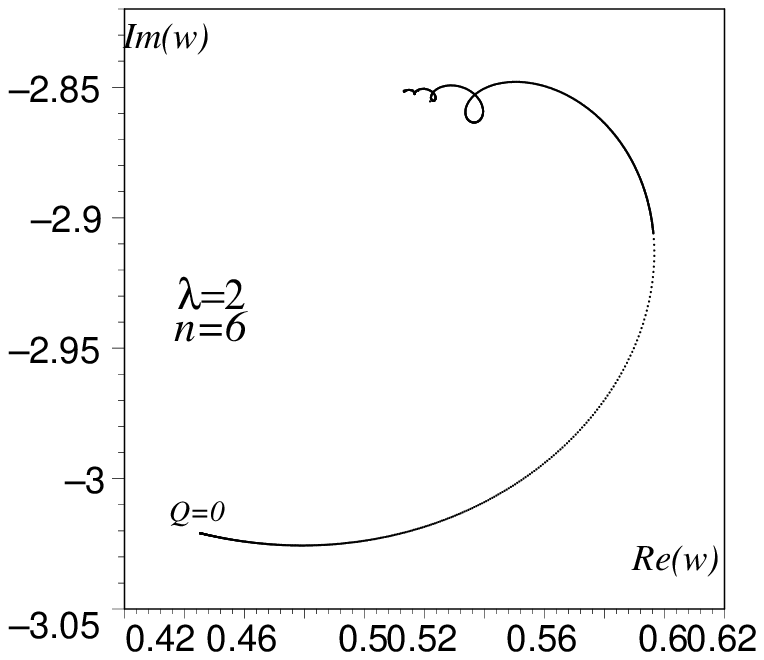} \hspace*{0.5cm}
\includegraphics[scale=0.6]{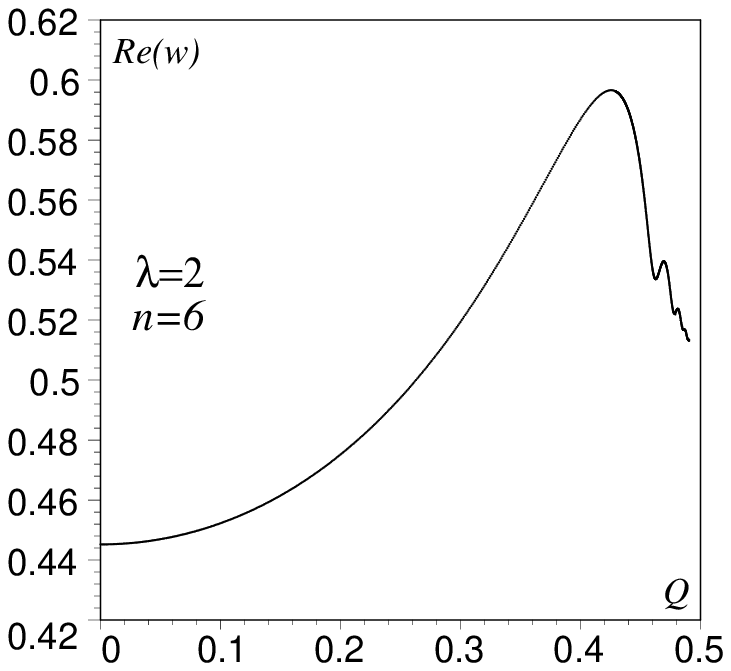} \hspace*{0.5cm}
\includegraphics[scale=0.6]{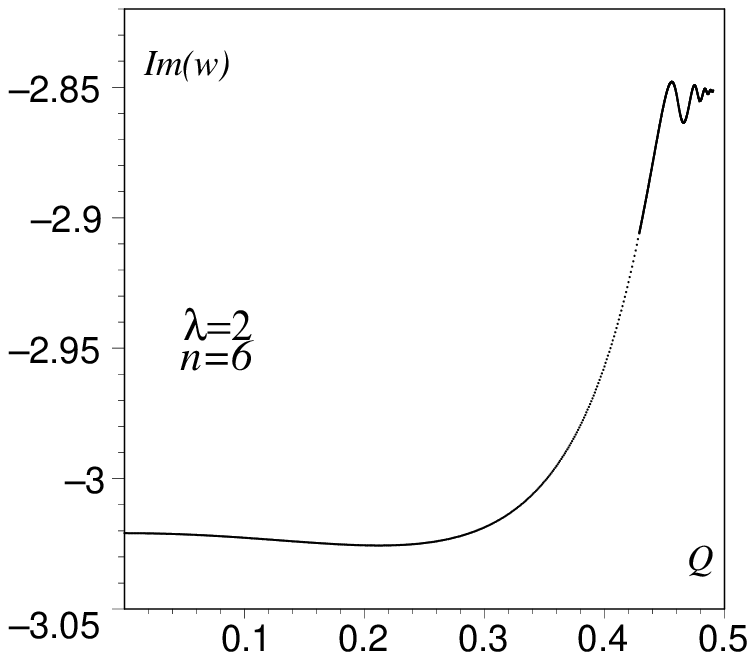}\vspace*{0.2cm} \\
\includegraphics[scale=0.6]{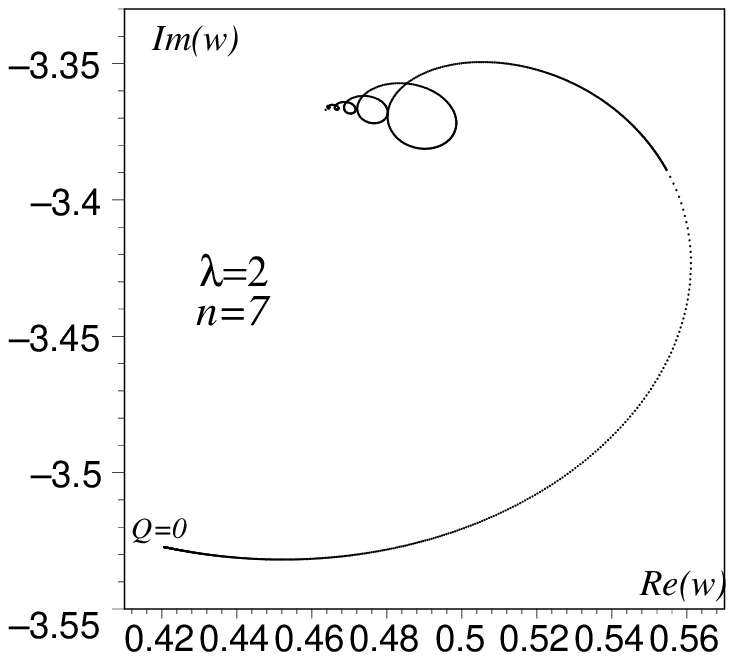} \hspace*{0.5cm}
\includegraphics[scale=0.6]{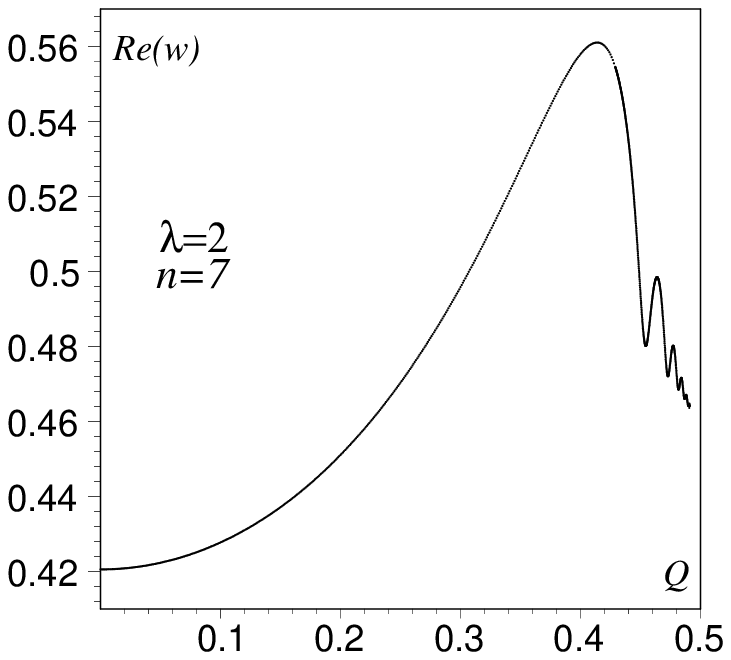} \hspace*{0.5cm}
\includegraphics[scale=0.6]{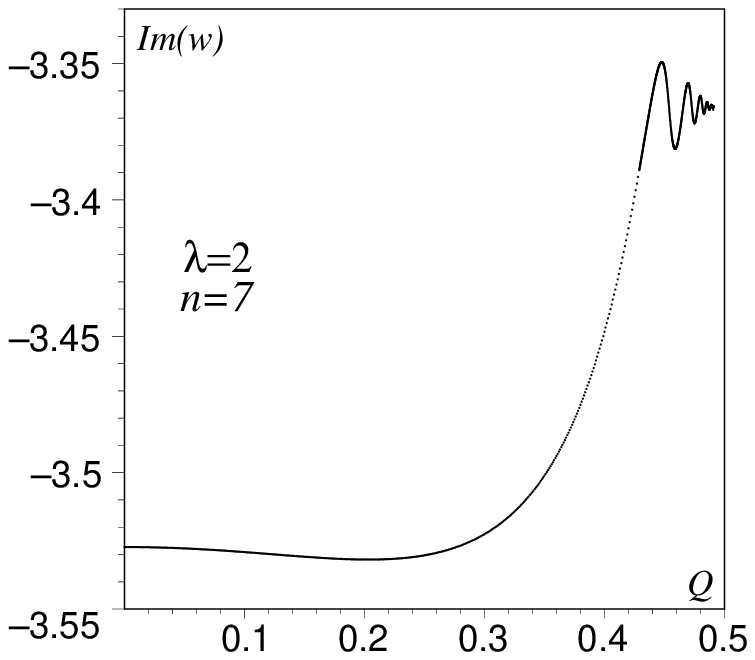}  \vspace*{0.2cm} \\
\includegraphics[scale=0.6]{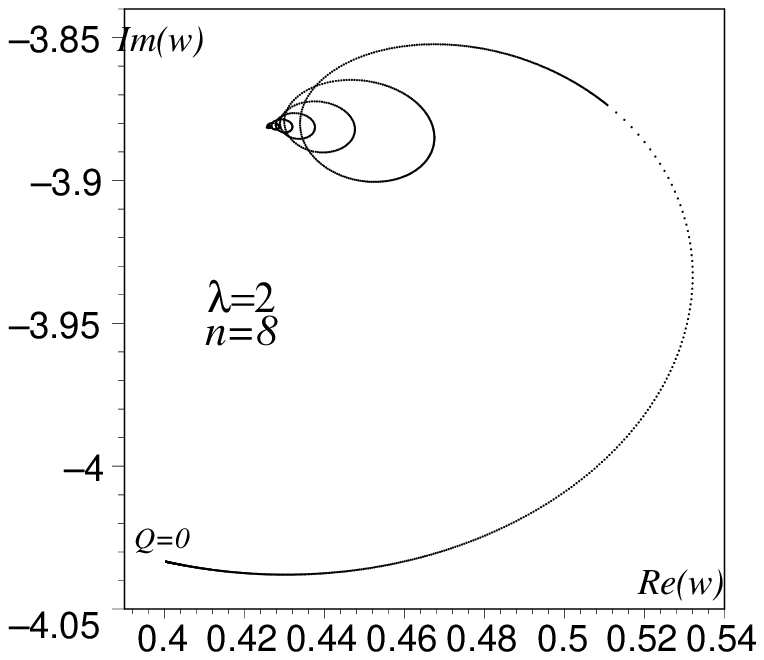} \hspace*{0.5cm}
\includegraphics[scale=0.6]{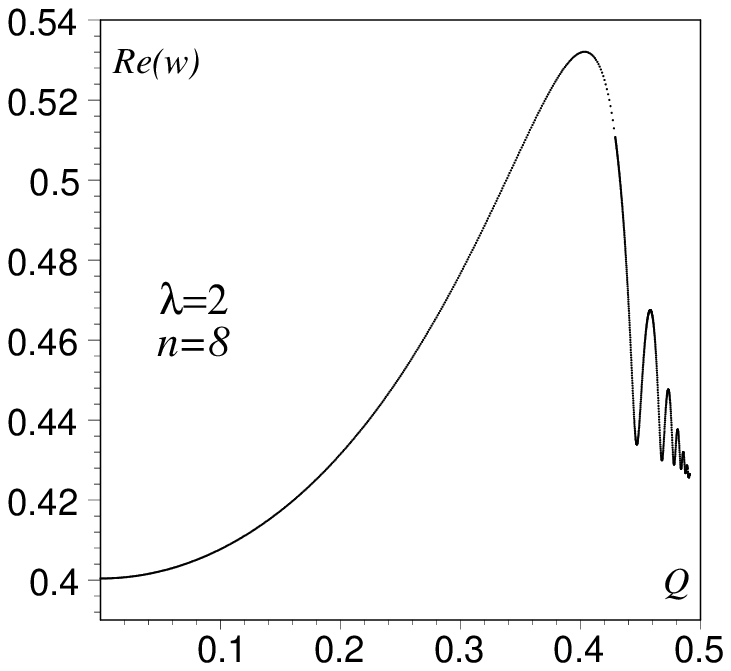} \hspace*{0.5cm}
\includegraphics[scale=0.6]{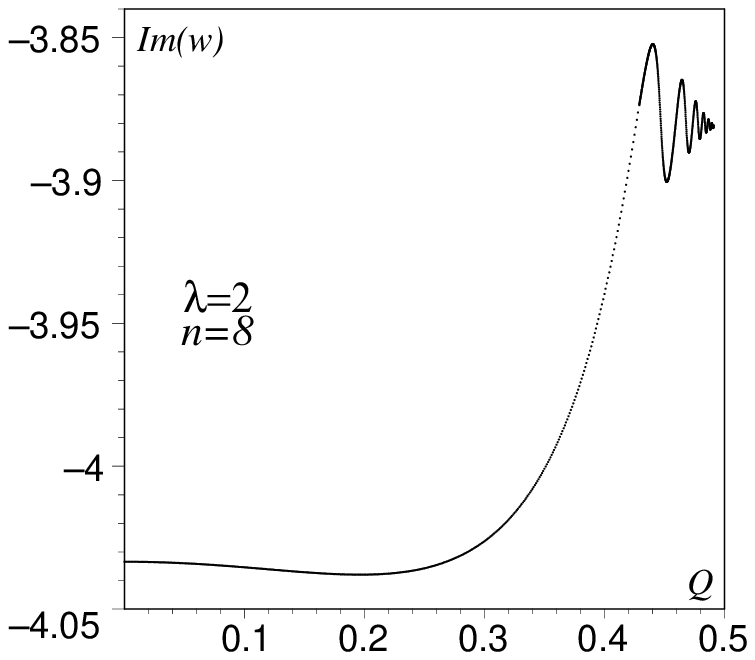}\vspace*{0.2cm} \\
\includegraphics[scale=0.6]{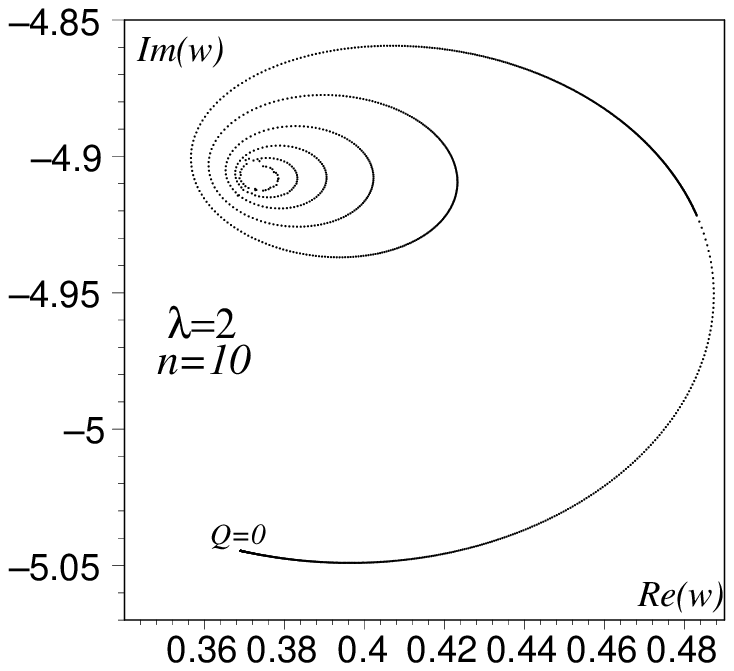} \hspace*{0.5cm}
\includegraphics[scale=0.6]{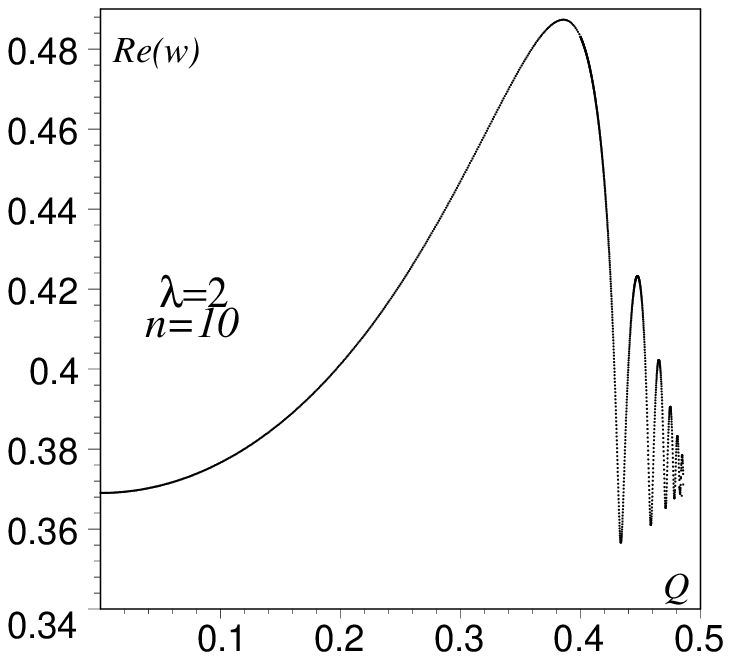} \hspace*{0.5cm}
\includegraphics[scale=0.6]{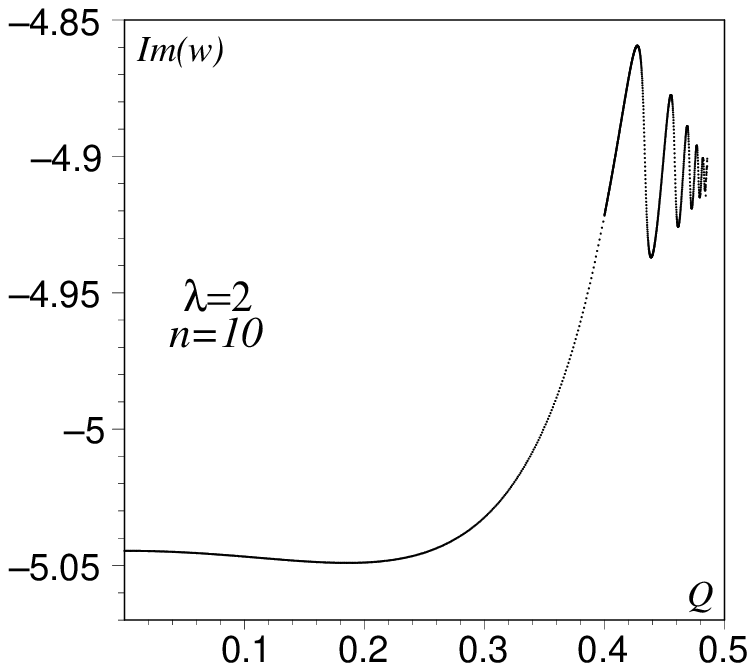}
\caption{\label{fig3} The left 5 figures describe the behavior of
quasinormal frequencies for $n=5,6,7,8,10$ and $\lambda=2$ in the
complex $\omega$ plane and the others graph $Re(\omega)$ and
$Im(\omega)$ of the quasinormal frequencies versus charge $Q$. By
comparing it with Fig. \ref{fig2} we know that the number of the
spirals decreases and the oscillations become slower as the
angular quantum number increases.}}

\section{summary}

The wave equation for the neutrino fields in the RN black hole
spacetime is obtained by means of the Newman-Penrose formulism.
The expansion coefficients of the wave equation satisfying
appropriate boundary conditions are determined by a three-term
recurrence relation. Then, the neutrino quasinormal frequencies of
the RN black hole spacetime are evaluated using continued fraction
approach and the results are presented by table and figures. We
find, for any angular quantum number, that the real part of the
fundamental quasinormal frequency increases as the charge
increases, but the imaginary part decreases first and then
increases as the charge increases. The fact tell us that the
intermediate decay of the Dirac perturbation around the RN black
hole depends on the charge. As the charge increases, the
oscillating frequency of this decay increases and the amplitude
per unit time increases first and then decreases. We then show
that the quasinormal frequencies become evenly spaced for large
$\lambda$ and the spacing of the real part is related to the
charge and that of the imaginary part becomes zero. We also find
that the frequencies in the complex $\omega$ plane generally move
counterclockwise as the charge is increased. They get a
spiral-like shape, moving out of their Schwarzschild value and
``looping in" towards some limiting frequency as the charge tends
to the extremal value.  For a given $\lambda$, we observe that the
number of spirals increases as the overtone number increases.
However, for a given overtone number, increasing $\lambda$  has
the effect of ``unwinding" the spirals. At last, we find that the
real and imaginary parts of the quasinormal frequencies are
oscillatory functions of charge, and the oscillating behavior
starts earlier and earlier as the overtone number grows for a
fixed $\lambda$, but it begins later and later as $\lambda$
increases for a fixed overtone number.

\begin{acknowledgments}This work was supported by the
National Natural Science Foundation of China under Grant No.
10473004; the FANEDD under Grant No. 200317; the SRFDP under Grant
No. 20040542003; the Hunan Provincial Natural Science Foundation
of China under Grant No. 05JJ0001.
\end{acknowledgments}

\end{document}